\documentclass[aps,prl,twocolumn,superscriptaddress,showpacs]{revtex4-1}

\usepackage{graphicx}
\usepackage{setspace}
\usepackage{array}
\newcolumntype{C}[1]{>{\centering\let\newline\\\arraybackslash\hspace{0pt}}m{#1}}

\begin{document}

\title{Very large domain wall velocities in Pt/Co/Gd trilayers with Dzyaloshinskii-Moriya interaction}

\
\author{Thai~Ha~Pham}
\affiliation{CNRS, Institut N\'{e}el, 38042 Grenoble, France}
\affiliation{Univ.~Grenoble Alpes, Institut N\'{e}el, 38042 Grenoble, France}
\
\author{J.~Vogel}
\affiliation{CNRS, Institut N\'{e}el, 38042 Grenoble, France}
\affiliation{Univ.~Grenoble Alpes, Institut N\'{e}el, 38042 Grenoble, France}
\
\author{J.~Sampaio}
\affiliation{Laboratoire de Physique des Solides, Univ. Paris-Sud, CNRS UMR 8502, 91405 Orsay, France}
\
\author{M.~Va\v{n}atka}
\affiliation{CNRS, Institut N\'{e}el, 38042 Grenoble, France}
\affiliation{Univ.~Grenoble Alpes, Institut N\'{e}el, 38042 Grenoble, France}
\
\author{J.-C.~Rojas }
\affiliation{CNRS, Institut N\'{e}el, 38042 Grenoble, France}
\affiliation{Univ.~Grenoble Alpes, Institut N\'{e}el, 38042 Grenoble, France}
\
\author{M. Bonfim}
\affiliation{Departamento de Engenharia El\'{e}trica, Universidade Federal do Paran\'{a}, Curitiba, Brazil}
\
\author{D.~S.~Chaves}
\affiliation{CNRS, Institut N\'{e}el, 38042 Grenoble, France}
\affiliation{Univ.~Grenoble Alpes, Institut N\'{e}el, 38042 Grenoble, France}
\
\author{F.~Choueikani}
\affiliation{Synchrotron Soleil, L'Orme des Merisiers, Saint-Aubin, 91192 Gif-sur-Yvette, France}
\
\author{P.~Ohresser}
\affiliation{Synchrotron Soleil, L'Orme des Merisiers, Saint-Aubin, 91192 Gif-sur-Yvette, France}
\
\author{E.~Otero}
\affiliation{Synchrotron Soleil, L'Orme des Merisiers, Saint-Aubin, 91192 Gif-sur-Yvette, France}
\
\author{A.~Thiaville}
\affiliation{Laboratoire de Physique des Solides, Univ. Paris-Sud, CNRS UMR 8502, 91405 Orsay, France}

\author{S.~Pizzini} \email{stefania.pizzini@neel.cnrs.fr}
\affiliation{CNRS, Institut N\'{e}el, 38042 Grenoble, France}
\affiliation{Univ.~Grenoble Alpes, Institut N\'{e}el, 38042 Grenoble, France}


\date{\today}

\begin{abstract}
We carried out measurements of domain wall (DW) velocities driven by magnetic field pulses in symmetric Pt/Co/Pt and asymmetric Pt/Co/AlOx, Pt/Co/GdOx and Pt/Co/Gd trilayers with ultrathin Co layers and perpendicular magnetic anisotropy. In agreement with theoretical models, the maximum observed velocity is much larger in the asymmetric samples, where the interfacial Dzyaloshinskii-Moriya interaction (DMI) stabilises chiral N\'{e}el walls, than in the symmetric stack. In addition, in Pt/Co/Gd very large DW speeds (up to 600 m/s) are obtained, 2.5 times larger than in samples with oxidised Gd. Magnetic measurements reveal that this may be explained by the anti-parallel coupling between the magnetic moments of  Gd and Co at the Gd/Co interface, leading to a decrease of the total magnetisation. In quantitative agreement with analytical models, in all samples the maximum observed DW speed scales as $D / M_\mathrm{s}$, where $D$ is the strength of the DMI and $M_\mathrm{s}$ the spontaneous magnetisation.
\end{abstract}

\maketitle

\section{Introduction}
Domain wall (DW) dynamics in multilayer films with perpendicular magnetic anisotropy (PMA) is an important topic today, as these micromagnetic objects may be used as carriers of binary information in future ultra-high density storage devices \cite{Parkin2008}. Recently, it was discovered that in non centrosymmetric multilayers with ultrathin magnetic films  DWs can acquire a fixed chirality stabilised by the interfacial Dzyaloshinskii-Moriya interaction (DMI). This has boosted the interest for such systems as this new feature strongly modifies the DW  dynamics, whether driven by field or by spin polarised current pulses. In particular, in systems with strong DMI, DWs move at  large speeds when driven by current pulses \cite{Moore2008,Miron2011,Ryu2013,Emori2013}. This effect is intrinsically related to the chiral N\'{e}el internal structure of the DW imposed by the presence of DMI \cite{Thiaville2012}.

While recent works have studied field-driven DW dynamics in the very low velocity regime \cite{Je2013,Hrabec2014,Lavrijsen2015,Vanatka2015}, few studies have explored the effect of the DMI on the high velocity regime of DW dynamics \cite{Yoshimura2015, Yamada2015, Jue2016}. The maximum DW velocity is limited by the breakdown occurring at the so-called Walker field $H_\mathrm{W}$, beyond which the DW motion is no longer stationary and its internal structure undergoes continous precession. In symmetric ultrathin films with PMA this breakdown is reached for small fields (around 10-20~mT for Pt/Co/Pt samples \cite{Metaxas2007}), as the energy difference between N\'{e}el and Bloch wall configurations is small and it is thus easy to switch from one to the other. For this reason the velocities observed at high fields are relatively small (60~m/s for 150~mT in \cite{Metaxas2007}).

This limitation is overcome in multilayer systems with strong interfacial DMI \cite{Dzyaloshinskii1957,Moriya1960}. This interaction, an antisymmetric exchange term favouring noncollinear magnetic textures, can be non-vanishing in asymmetric stacks in which an ultrathin magnetic film is deposited on a large spin-orbit metal layer like Pt, Ta or W \cite{Fert1990}. The DMI can be seen as a chiral field $H_\mathrm{DMI}$ normal to the DW direction and localised on the DW, which induces homochiral N\'{e}el walls that become stable against precession up to much larger fields \cite{Thiaville2012}. This allows them to reach much larger velocities than in symmetric stacks without DMI. A first demonstration was given by Miron \textit{et al.} \cite{Moore2008} who showed that in Pt/Co/AlOx trilayers DWs move with a larger velocity than in Pt/Co/Pt. The presence of a large DMI in Pt/Co/AlOx was confirmed experimentally using chiral nucleation experiments \cite{Pizzini2014} and Brillouin Light Scattering spectroscopy (BLS) \cite{Belmeguenai2015}.

The aim of this work is to compare DW velocities in symmetric and asymmetric stacks with PMA and to address the effect of the DMI strength on the DW velocity at the Walker field and above. We show that velocities much larger than those measured for symmetric Pt/Co/Pt can be obtained in asymmetric Pt/Co/MOx (M=Al, Gd), and that these increase considerably in samples where Co and Gd interfacial moments couple antiparallel. We also show that in the asymmetric samples the velocity saturates for large applied out-of-plane fields, in contrast with 1D micromagnetic simulations that predict a drop of the speed at the Walker field, but in agreement with 2D simulations.

\section{Sample growth and characterisations}
Measurements were carried out on a Pt(4)/Co(1)/Pt(4) symmetric stack and three asymmetric stacks, Pt(4)/Co(0.8)/AlOx(3), Pt(4)/Co(1)/GdOx(4)/Al(7) and Pt(4)/Co(1)/Gd(3)/Al(7) (thickness in nm). All samples were grown on Si/SiO$_{2}$ substrates by magnetron sputtering at room temperature and the Al and Gd layers were oxidised with an oxygen plasma. Note that the Co thicknesses are larger than the 0.6~nm value of previous works \cite{Miron2011,Jue2016}, so as to bring the Walker breakdown into the observable field region. The magnetic properties of the samples were obtained from VSM-SQUID measurements at variable temperature and from X-ray Magnetic Circular Dichroism (XMCD) measurements at the Co $L_{2,3}$ and Gd $M_{4,5}$ absorption edges. The XMCD measurements were carried out at the DEIMOS beamline of the SOLEIL synchrotron (Saint Aubin, France).

The DW velocities were measured on continuous films, using wide field magneto-optical Kerr microscopy. The film magnetisation was first saturated in the out-of-plane direction. An opposite  magnetic field pulse was then applied to nucleate one or several reverse domains. The DW velocity was deduced from the expansion of the initial bubble domain, after the application of further magnetic field pulses. Using a 200~$\mu$m diameter coil associated to a fast current source, out-of-plane magnetic field pulses of strength up to $B_\mathrm{z}$=300 mT and duration down to 20~ns could be applied. Two kinds of measurements were carried out. In order to visualize the various regimes of DW propagation predicted by the theoretical models, DW speeds were measured as a function of $H_\mathrm{z}$. In order to address the DMI strength in the various samples, the DW speeds driven by the $H_\mathrm{z}$ pulses were measured in the presence of a static in-plane field $H_\mathrm{x}$ normal to the DW plane
\cite{Je2013,Hrabec2014,Lavrijsen2015,Vanatka2015}.

\section{Domain wall dynamics in symmetric and asymmetric trilayers with PMA}
\begin{figure}
\includegraphics[width=8.8cm]{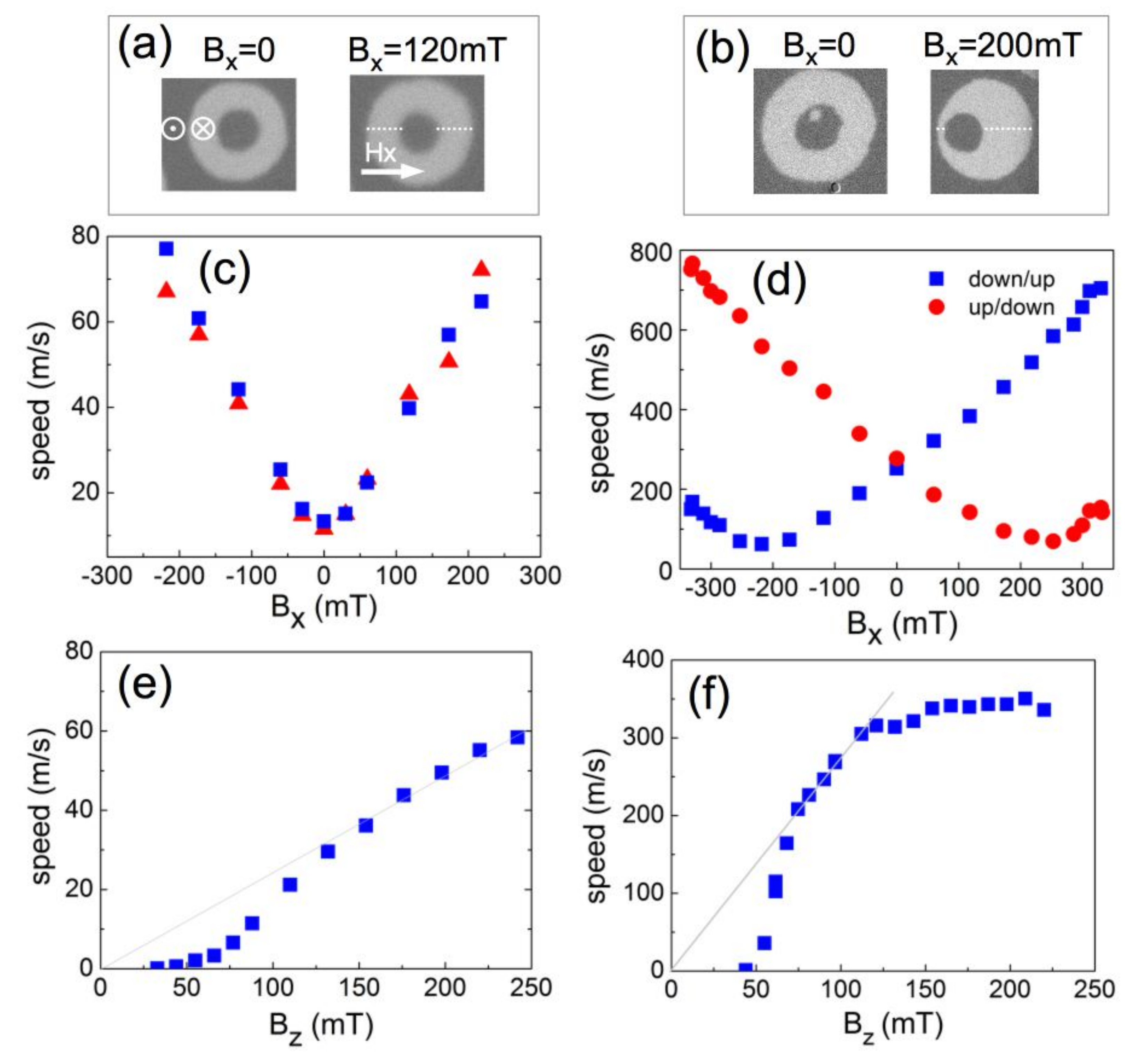}
\caption{(a)-(b): differential Kerr images showing the expansion of a domain during the application of an out-of-plane field $B_\mathrm{z}$, without and with the simultaneous application of an in-plane field, in Pt/Co/Pt (a) and Pt/Co/AlOx (b); (c)-(d): DW velocity \textit{vs.} in-plane field $B_\mathrm{x}$ for Pt/Co/Pt ($B_\mathrm{z}$ = 88 mT) (c) and Pt/Co/AlOx ($B_\mathrm{z}$ = 132 mT) (d). (e)-(f): DW velocity \textit{vs} $B_{z}$ for Pt/Co/Pt (e) and Pt/Co/AlOx (f). The thin lines in (e,f) emphasize the slope of the linear regime.}
\label{fig:Figure1}
\end{figure}

Fig.~\ref{fig:Figure1}(a-b) shows the differential Kerr images corresponding to the expansion of a domain in Pt/Co/Pt (a) and Pt/Co/AlOx (b) after the application of a $H_\mathrm{z}$ pulse, respectively without in-plane field (left images) and in the presence of a strong in plane field (right images). In Fig.~\ref{fig:Figure1}(c-d) we show the DW speed~\textit{vs.}~$B_\mathrm{x}$ curves measured for the up/down and down/up DWs propagating in the $\mp$x directions, driven by 20~ns long $B_\mathrm{z}$ field pulses of respectively 88~mT and 132~mT strength. For Pt/Co/Pt the speeds of up/down and down/up DWs are, within the error bars, the same for every value of $H_\mathrm{x}$. The curves for the two DWs are symmetric with respect to the $H_\mathrm{x}$~=~0 axis.

In general in symmetric magnetic layers with PMA, Bloch walls are more stable as they give rise to no magnetostatic energy while N\'{e}el walls do. In order to decrease their Zeeman energy, the magnetisation of such  walls align parallel to an in plane field applied perpendicular to them, as soon as this field is larger than the DW demagnetising field. In the high field regime, the DW velocity increases, as the DW width increases. The symmetric curves measured for Pt/Co/Pt reflect the equivalent dynamics of up/down and down/up DWs, and are in line with the presence in this sample of (non chiral) Bloch walls. Note that deviations from this symmetric trend have been observed by several groups in the low velocity regime \cite{Lavrijsen2015,Jue2015a,Vanatka2015}.

In Pt/Co/AlOx the DW propagation is anisotropic in the direction of $H_\mathrm{x}$. The speed~\textit{vs.}~$B_\mathrm{x}$ curves show a minimum for $\mp$~220mT respectively for down/up and up/down DWs, and the trend for up/down and down/up DWs is similar for $H_\mathrm{x}$ fields applied in opposite directions. This is the fingerprint of chiral N\'{e}el walls, where the Dzyaloshinskii-Moriya interaction acts as a longitudinal chiral field $H_\mathrm{DMI}$ localised on the DWs, with opposite directions for up/down and down/up DWs. The DW velocity presents a minimum when the applied in-plane field compensates the $H_\mathrm{DMI}$ field \textit{i.e.} for the $H_\mathrm{x}$ field for which the DW magnetization reaches the Bloch orientation \cite{Je2013}. From the value of this field we can then deduce the average DMI energy density $D$, since $H_\mathrm{DMI} = D /(\mu_{0}M_{s} \Delta )$ (where $D$ is the DMI strength, $M_\mathrm{s}$ the spontaneous magnetisation and $\Delta = \sqrt{A/K_{0}}$  the DW width parameter). The large DMI energy density ($D$~=~1.63~mJ/m$^2$ for 0.8~nm Co thickness) - extracted using $\mu_{0}H_\mathrm{DMI}$~=~220~mT and the material parameters shown in Table 1 - is in good agreement with the values previously  obtained for similar samples  \cite{Pizzini2014,Belmeguenai2015}. The sign of the $H_{DMI}$ field confirms the presence of left handed chirality \cite{Pizzini2014,Jue2016,Tetienne2014}.

Let us now see what is the consequence of the different DW structure on the mobility of the DWs driven by a pure easy-axis field. Figs.~\ref{fig:Figure1}(e-f) show the DW velocity as a function of $B_\mathrm{z}$, measured for the Pt/Co/Pt and Pt/Co/AlOx trilayers. The trend  for Pt/Co/Pt is very similar to that found by \cite{Metaxas2007} for  Pt/Co(0.8nm)/Pt, where the velocity reached a maximum value of 70~m/s for $B_\mathrm{z}$=140~mT. In our Pt/Co/Pt the velocity changes exponentially with the field up to around 130~mT ; beyond this field the DW mobility is constant up to the largest applied field (250~mT). According to the measured material parameters, the Walker field in this sample ($H_\mathrm{W}=\alpha M_{S}N_\mathrm{NDW}$/2 where $N_\mathrm{NDW}$ is the demagnetisation factor of a N\'{e}el wall) should be of the order of 12.5~mT and is therefore hidden by the creep regime. We therefore expect that in the constant mobility region the DWs evolve in the so-called precessional flow, where the DW internal structure continuously changes between Bloch to N\'{e}el. In this regime the asymptotic mobility $m=\gamma_{0}\Delta/(\alpha+1/\alpha)$ = 0.25 m/s/mT is relatively small and  in good agreement with that obtained in \cite{Metaxas2007}. From a linear fit of the high field curve we obtain $\alpha\approx0.4$. On the other hand, fitting the data under the assumption that the linear regime is associated to the steady flow ($m=\gamma_{0}\Delta/\alpha$) gives rise to an irrealistic value of the damping parameter ($\alpha=4$).

The speed~\textit{vs.}~$B_\mathrm{z}$ curve measured for Pt/Co/AlOx strongly differs from that measured for Pt/Co/Pt, with the striking result that the maximum DW velocity in the given field range is a factor 5 larger. The depinning of the DWs occurs for $\approx$60~mT so that for fields between 70 and 120~mT a clear linear regime appears, characterised by a DW mobility of 3m/s/mT. We correlate the large velocities with the strong DMI evidenced by the previous measurements. Thiaville \textit{et al.} \cite{Thiaville2012} predict that in the presence of large DMI the Walker breakdown shifts to larger fields and that $H_{W}\approx \alpha H_\mathrm{D}= \pi\alpha H_\mathrm{DMI}/2 $. Using $\alpha\approx 0.35$, as extracted from the fit of the DW mobility in the linear regime and the measured $\mu_{0}H_\mathrm{DMI}$, we obtain $\mu_{0}H_\mathrm{W}\approx$ 120~mT which is, as predicted, much larger than that of Pt/Co/Pt, and very close to the field for which the linear regime ends and the DW velocity practically saturates in our measurements. Note also that in agreement with the prediction of 1D micromagnetic models \cite{Thiaville2012} the Walker velocity changes linearly with the in-plane field. The  inset of Fig.~\ref{fig:Figure3}(b) shows that in the Pt/Co/Gd trilayer described below, the velocity reaches 1100m/s for $B_\mathrm{x}$=300mT.  The DMI-induced extension of the steady flow regime therefore allows explaining the large DW velocities observed for the asymmetric samples.

\section{Saturation of DW speeds at large fields}
\begin{figure}
\includegraphics[width=8.2cm]{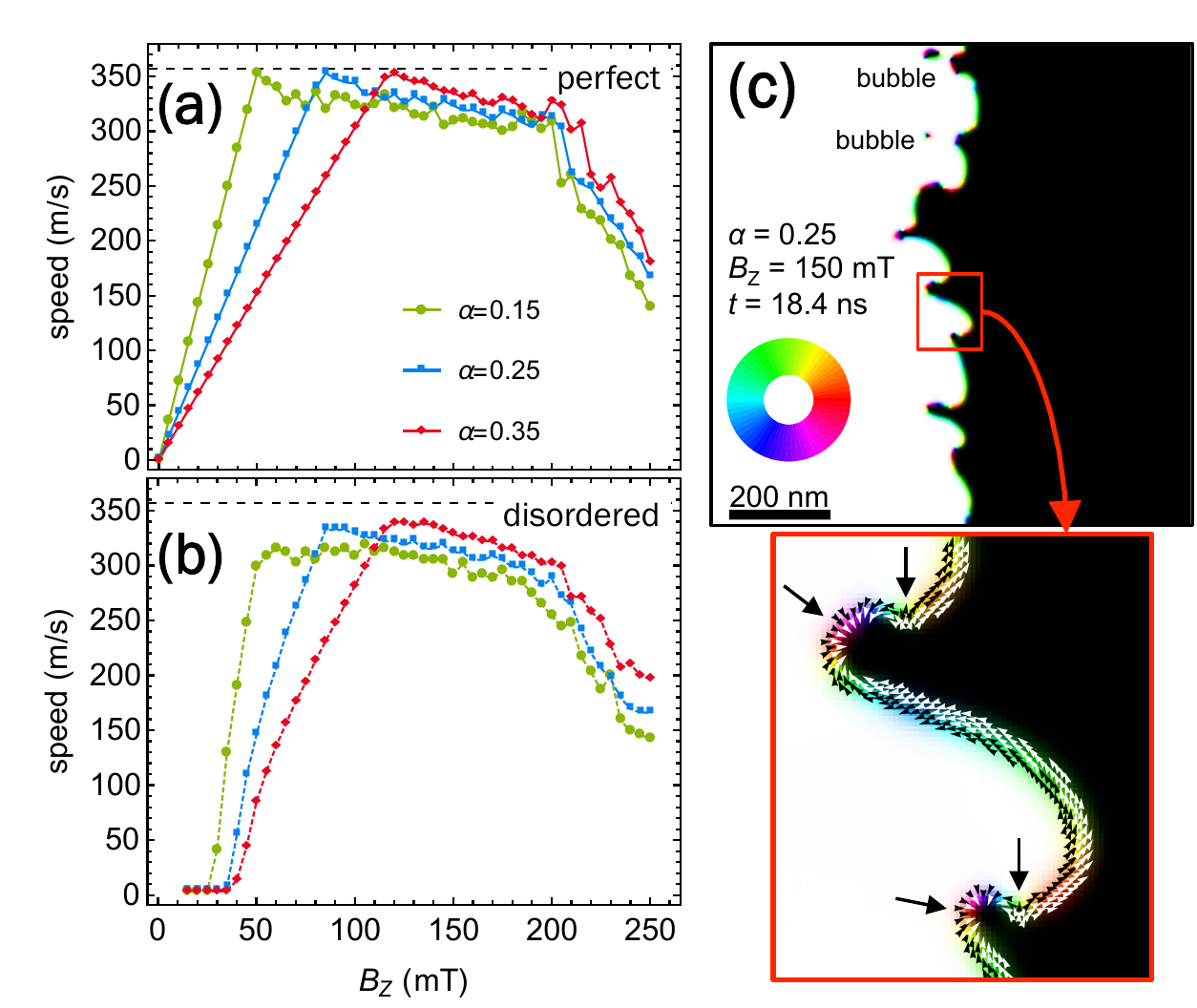}
\caption{Simulated DW dynamics in a 1~$\mu$m~x~1~$\mu$m moving window, with the parameters of Pt/Co/GdOx: $A$ = 16 pJ/m, $M_\mathrm{s}$ = 1.26MA/m, $D$ = 1.5 mJ.m$^{-2}$, $K_\mathrm{u}$ = 1.44 MJ.m$^{-3}$. A perfect sample (a,c) and a disordered sample (b) are considered. In the magnetisation snapshot (c), white/black corresponds to the perpendicular magnetisation component and the colours to the in-plane component (see inset colour wheel). Large arrows indicate the Bloch lines.}
\label{fig:Figure2}
\end{figure}

In contrast with the predictions of micromagnetic 1D models, in Pt/Co/AlOx, as well as in Pt/Co/Gd and Pt/Co/GdOx shown below, the DW velocity saturates for large $H_\mathrm{z}$ fields. We carried out large scale 2D micromagnetic simulations using the MuMax3 code \cite{Mumax3} with the parameters of Pt/Co/GdOx determined in this paper (see Table~1). To avoid the effects of the borders and the associated DW tilting \cite{Yoshimura2015,Jue2016} we simulated a propagating DW in a moving 1~x~1~$\mu$m$^2$~x~~1nm box with periodic boundary conditions linking the top and bottom edges. The lateral mesh size was 2~nm. Different values of the damping $\alpha$ (0.15 to 0.35) were tested. In addition, perfect and disordered samples were considered (the latter realised with a spatial fluctuation of the uniaxial anisotropy value, as was done in \cite{Jue2016}).
Fig.~\ref{fig:Figure2}(a) shows the results for the perfect system. The DW speed initially increases linearly with field, up to the Walker field, which is proportional to the damping parameter. The Walker velocity is independent of damping, as expected from the 1D collective coordinates (q,$\Phi$) model \cite{Thiaville2012}. Above the Walker field, the speed reaches a plateau with $\approx 300$m/s, in quantitative agreement with the experimental results (Fig.~\ref{fig:Figure3}(a)). This plateau is a characteristic of the dissipation mechanisms in 2D DMI DWs above the Walker field \cite{Yoshimura2015,Yamada2015}. A representative snapshot of the DW structure above the Walker field is shown in Fig.~\ref{fig:Figure2}(c). The DW presents a complex meander shape and its magnetisation rotates several times along the wall. As discussed by Yoshimura et al. \cite{Yoshimura2015}, pairs of Bloch lines (marked with large arrows in the magnified view) are continuously nucleated in the DW. These lines, with their N\'{e}el-like orientation, can have two very different widths, depending on whether they rotate in the sense favoured by DMI or not. The DW sections containing lines are slower, hence the meandering DW shape. The lines disappear either by collapse of winding pairs with emission of spin waves (like in \cite{Yoshimura2015}), or by the creation of bubble domains (visible in the figure) that detach by pinching off the wall meanders, and eventually collapse. When the bubbles contain Bloch lines, these annihilate and transform the bubbles into skyrmions with structure stabilised by DMI, before the skyrmions themselves collapse. The creation of bubbles is a second dissipation mechanism that was not seen in the calculations by Yoshimura \textit{et al.}, as the DMI considered in that work was relatively small (typically 16\% of the critical value $D_\mathrm{c}$ for destabilisation of the uniform magnetisation), whereas here $D/D_\mathrm{c}$= 0.45. The  reduced DW energy in the case considered here favours the meandering of the DW and the subsequent creation of bubbles. Altogether, and as explained by \cite{Yoshimura2015} who called it a solitonic behaviour, the DW structure is on the average that of a chiral Bloch wall. Indeed, with dominant DMI, this is the DW structure just before the Walker breakdown \cite{Thiaville2012}. In the case with disorder, the effects are qualitatively the same, as seen in Fig.~\ref{fig:Figure2}(b). As expected, the disorder induces a non-zero propagation field, but does not change the Walker field nor the maximum velocity. The shape of the DW and the energy dissipation mechanism (collapse of Bloch line pairs and creation of detached bubbles) is also not significantly affected.
At fields beyond the experiment range (about 200~mT), the simulations reveal that the plateau ends and the DW velocity decreases. This indicates that the dissipation mechanisms can no longer sustain a constant velocity. In fact, the data show that the dissipated power reduces.

\section{Influence of the saturation magnetisation on the maximum DW speed}

\begin{figure}
\includegraphics[width=8.8cm]{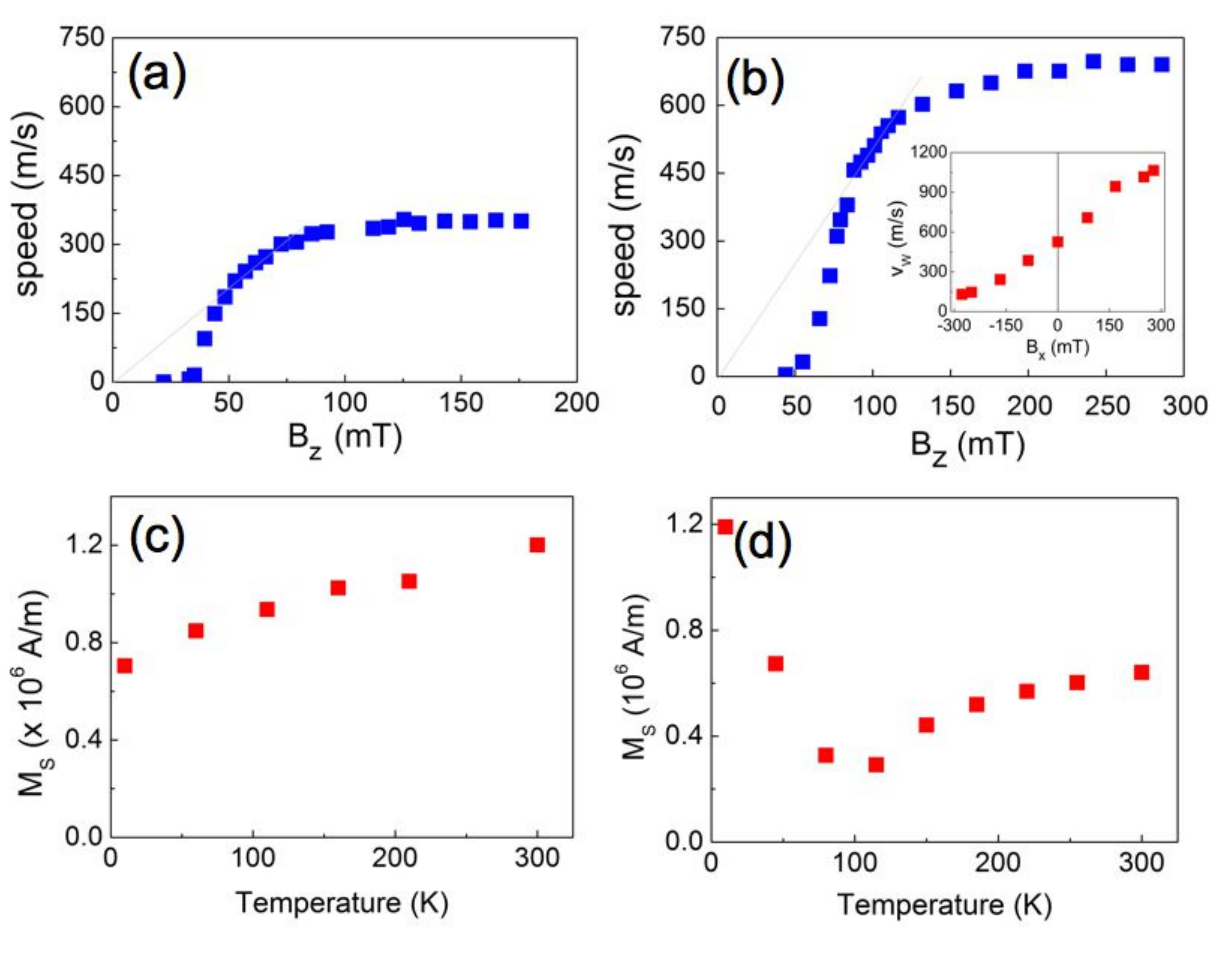}
\caption{(a-b) DW speed \textit{vs.}~$H_\mathrm{z}$ field measured for  Pt/Co/GdOx (a) and  Pt/Co/Gd  (b) Inset of (b): Walker velocity as a function of $H_\mathrm{x}$. The thin lines emphasise the slope of the linear regime.  (c-d)  magnetisation \textit{vs.} temperature normalised to a 1~nm thickness, measured by VSM-SQUID for Pt/Co/GdOx (c) and Pt/Co/Gd (d).  }
\label{fig:Figure3}
\end{figure}

In order to address the role of the top interface on the strength of the DMI and as a consequence on the maximum attainable DW velocity, we have studied samples in which the top AlOx layer was replaced by GdOx or Gd. Figure~\ref{fig:Figure3}(a,b) shows the DW speed \textit{vs.}~$B_\mathrm{z}$ field measured for two samples: Pt/Co/GdOx with a completely oxidised Gd layer and a Pt/Co/Gd sample where the Gd layer was not deliberately oxidised. The most striking result is that while the saturation DW speed in the Pt/Co/GdOx is of the same order of magnitude as that measured for Pt/Co/AlOx (around 300~m/s), this velocity in Pt/Co/Gd is a factor 2 larger.

In order to clarify this behaviour, we examined the magnetic properties of the two samples. Both samples have out-of-plane magnetic anisotropy with very similar anisotropy fields. VSM-SQUID measurements (Fig.~\ref{fig:Figure3}(c-d)) show that the magnetisation of Pt/Co/GdOx changes weakly when going from 300~K to 10~K : since the moment of Co is expected to change little in this temperature range ($T_\mathrm{C}\gg300$~K), the slight decrease of the magnetisation can be attributed to the presence in the sample of a small amount of non-oxidised Gd, which may be intermixed with the Co and acquire a moment antiparallel to that of Co (we evaluate this amount of Gd to less than 0.1 atomic layer). On the other hand in Pt/Co/Gd the magnetisation is strongly temperature dependent, with an increase of the magnetisation at low temperature and a minimum at around 80-100~K. This is the behaviour found for ferrimagnetic rare-earth/transition metal (RE-TM) compounds, where the RE moment strongly increases at low temperature while the TM moment changes little. This results in the presence of a compensation temperature, where the two sublattice magnetisations are equal. From this behaviour we can be confident that a large part of the Gd layer is non-oxidised in this sample, the Co and Gd layers are coupled at the interface and their moments align antiparallel to each other. Note also that while the magnetisation measured for Pt/Co/GdOx is $12.6\times10^{5}$~A/m, that of Pt/Co/Gd is $6.4\times10^{5}$~A/m at room temperature (RT), assuming a 1~nm magnetic layer thickness. This is a clear indication that the Gd contributes to the total magnetisation even at RT.

\begin{figure}
\includegraphics[width=8.8cm]{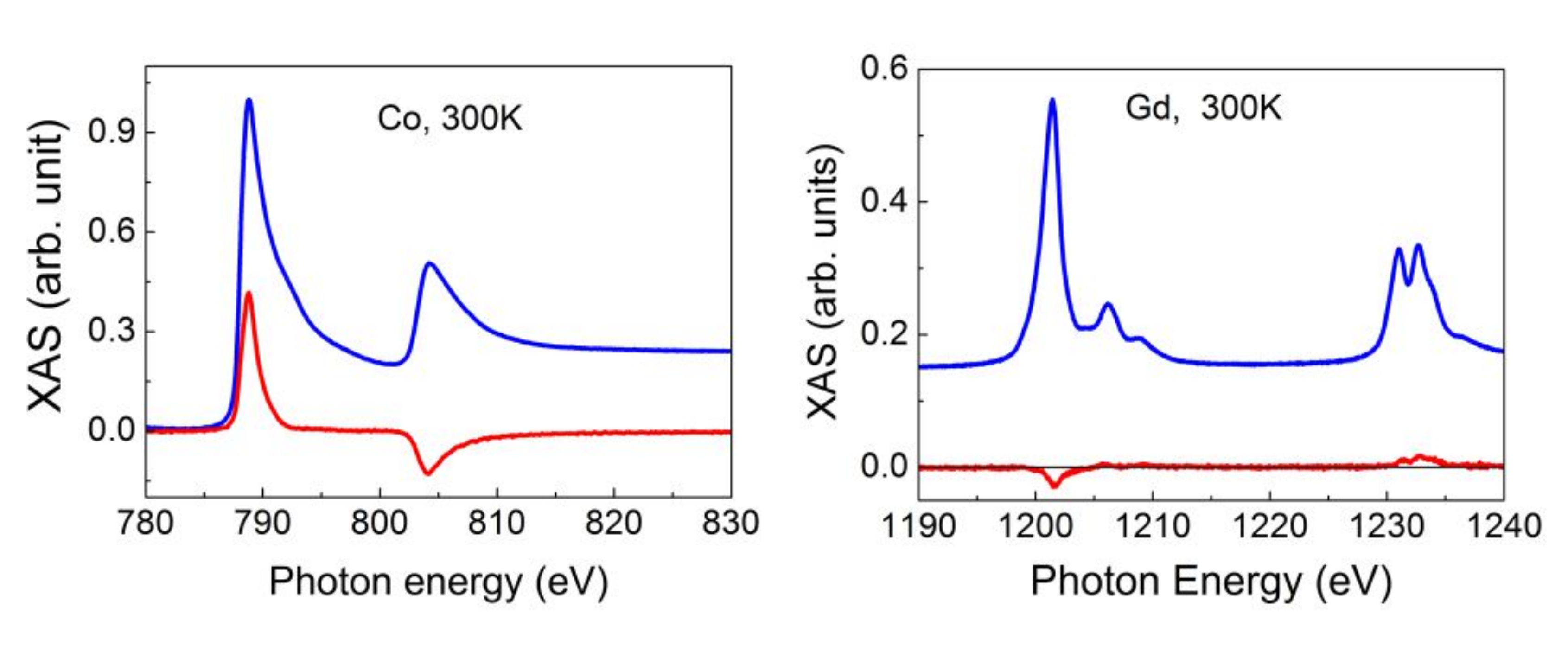}
\caption{XAS (blue line) and XMCD measurements (red line) of the Pt/Co/Gd sample at the Co $L_{2,3}$ and Gd $M_{4,5}$ edges, at 300~K and in an applied magnetic field of 1~T.}
\label{fig:Figure4}
\end{figure}

In order to quantify the contribution of Gd to the total magnetisation in the Pt/Co/Gd sample and at RT, we have carried out XMCD measurements at the Gd $M_{4,5}$ and Co $L_{2,3}$ edges. The X-ray Absorption and XMCD spectra measured at 300~K  in a magnetic field of 1~T are shown in Fig.~\ref{fig:Figure4}. The integrated XMCD signals can be related to the magnetic moment per absorbing atom \cite{Thole1992, Carra1993}. A non-vanishing XMCD signal with sign opposite to that of Co is found at the Gd edge at 300~K. This emphasises: i) that part of the Gd is magnetic at RT and ii) that the Gd $4f$ and Co $3d$ moments are antiparallel. At 4~K, the Co and Gd XMCD signs are reversed and the Gd signal is much larger (not shown). This is coherent with the fact that at low temperature the Gd magnetisation increases, so that the total Gd moment is larger than that of Co and aligns with the external magnetic field.

If we suppose that the Gd intermixed with the Co at the interface has a magnetic moment similar to the one measured by L\'{o}pez-Flores  \textit{et al.} \cite{Boeglin2013} at 290~K (4.5 $\mu_B$) in Gd$_\mathrm{20}$Co$_\mathrm{80}$, we can deduce that 10$\%$ (0.3 nm) of the Gd layer is magnetic at RT. This is the proportion of Gd atoms that is coupled to Co (either through exchange or through interdiffusion). In that case, we obtain that the contribution of Gd to the moment per unit surface found by VSM-SQUID ($6.4 \times 10^{-4}$ A) is $-4.1 \times 10^{-4}$ A, while the contribution of Co is $10.5 \times 10^{-4}$ A. For a Co thickness of 1~nm, this corresponds to M$_\mathrm{s} = 10.5 \times 10^{5}$ A/m, in reasonable agreement with the one obtained from XMCD (M$_S = 14 \times 10^{5}$ A/m) and that of Co found e.g. in the Pt/Co/GdOx sample.

\begin{table*}[t]
\caption{Co thickness  $t_{Co}$, spontaneous magnetisation $M_\mathrm{s}$, effective anisotropy energy $K_\mathrm{0}$, exchange stiffness $A$ (two values taken from literature), DW parameter $\Delta$, damping constant $\alpha$, DMI field $\mu_\mathrm{0}H_\mathrm{DMI}$, DMI energy density $D$ extracted from DMI field, DMI energy density $D^{*}$ obtained from other measurements:  a) nucleation \cite{Pizzini2014}, b) BLS, on a similar sample \cite{Vanatka2015}, c) BLS on the same sample. Note that the best agreement between $D$ and $D^{*}$ is obtained using $A$~=~16~pJ/m.}
\label{fig:Table1}
\scriptsize
\begin{center}
{\setstretch{1.15}
\begin{tabular}{ c   c c c  c  c  c c c  c  c  c  c  c }
\hline \hline
sample  & $t_{Co}$  & $M_\mathrm{s}$  &    $K_\mathrm{0}$ & $A$ & $\Delta$  &  $\alpha$ & $\mu_\mathrm{0}H_\mathrm{DMI}$  & $D$     & $D^{*}$ &  $\mu_\mathrm{0}H_\mathrm{W}^{th}$ &  $\mu_\mathrm{0}H_\mathrm{W}^{exp}$ \\\hline
        & nm   & MA/m  &     MJ/m$^{3}$ & pJ/m &  nm   &     & T  & mJ/m$^{2} $   &  mJ/m$^{2}$ &  mT &  mT   \\\hline
Pt/Co/AlOx & 0.8 & 1.18 &   0.41 & 22 & 7.2 &  0.45 & 0.22 & 1.91 &             & 140 &      \\
           &     &     &   &       16 & 6.2 &    0.39 &      & 1.63 & 1.65~$^{a}$ & 120 & 120 \\
Pt/Co/GdOx & 1   & 1.26 &  0.41 & 22 & 7.2 &  0.33 & 0.20 & 1.73 &             & 97 &     \\
           &     &      &  &       16 & 6.2      & 0.28 &     & 1.48 & 1.5~$^{b} $  & 83 & 79 \\
Pt/Co/Gd   & 1   & 0.64 &   0.26 & 22 & 9.2 &  0.34 & 0.30 & 1.78 &             & 156 &    \\
           &     &     &   &       16 & 7.9      & 0.30 &     & 1.52 & 1.5~$^{c}$   & 133 & 116\\
\hline
\hline
\end{tabular}}
\end{center}
\end{table*}

Let us now see what is the effect of the different nature of the Gd layer on the DW dynamics. The value of the  DW mobilities for Pt/Co/GdOx and Pt/Co/Gd may be obtained from the slope of the short linear regime appearing in the $v(B_\mathrm{z})$ curves, at the end of the depinning regime (see Fig.~\ref{fig:Figure3}) : these are respectively $\approx$ 4~m/s/mT and 5~m/s/mT. From these values and those of the effective magnetic anisotropy we deduce the damping parameters $\alpha$ (see Tab.~1), which we find to be similar for the two samples (and similar to the Pt/Co/AlOx sample.) This is expected as Gd has no orbital moment. Thus, the difference between the DW velocity found in the two samples does not come from the different damping.

We next look at the value of $D$, which can be  extracted from the measured $H_\mathrm{DMI}$ fields (column $D$ in Tab.~1). The value $D$ depends on the exchange stiffness $A$ (as it is the case for $\alpha$). This is the only material parameter that we could not measure. Using $A$ values between 16 and 22~pJ/m, we find $D$ values between 1.4 and 1.7~mJ/m$^2$. The best agreement with the $D$ values measured with BLS spectroscopy is obtained for both samples using $A = 16$~pJ/m (Tab.~1). The similarity of $D$ found for the two samples is an interesting result. Since the Pt/Co interfaces are very similar (they were grown in the same run) and hence contribute equally to the global DMI strength, our measurements suggest that the top Co interfaces may also give similar contributions to the global DMI. Considering the very different nature of the Co/GdOx and Co/Gd interfaces, we deduce that the top interface gives a negligible contribution to the total DMI.

Although we found the same values of $\alpha$ and $D$, the DW speed at the Walker field, as well as the terminal velocity in the plateau, is about twice larger in the Pt/Co/Gd than in the Pt/Co/GdOx sample. Following Thiaville \textit{et al.} for samples with strong DMI (i.e. $D >> 0.14 \mu_\mathrm{0} M_\mathrm{s}^2 t_\mathrm{Co}$) \cite{Thiaville2012}, the DW speed at the Walker field is given by:

   \begin{equation}v_{W} = \gamma_{0} \frac{\Delta}{\alpha} H_{W}\approx\frac{\pi}{2} \gamma_{0} \Delta H_{DMI} = \frac{\pi}{2} \gamma \frac{D}{M_{s}}
    \end{equation}

Therefore, $v_\mathrm{W}$ is related to the $H_\mathrm{DMI} \Delta$ product, and does not depend on $\alpha$. This is intuitive, as the DMI field increases the Walker field, and the widening of the DW increases the mobility. Indeed for the Pt/Co/Gd sample, a larger DMI field and larger DW width are found (Tab.~1). By inserting in Eq.~1 the experimental $H_\mathrm{DMI}$ and the calculated DW parameter, we obtain that within this model the ratio between the Walker velocities in Pt/Co/Gd and Pt/Co/GdOx is $\approx$~1.90. This is very close to the ratio 1.8 between the experimental values of the DW velocities at the end of the linear regime (570~m/s for Pt/Co/Gd and 320~m/s for Pt/Co/GdOx), and even closer to the ratio of the terminal velocities (1.9).

The second part of Eq. 1 shows that, very interestingly, v$_\mathrm{W}$ is proportional to $D/M_\mathrm{s}$, and is independent of exchange and damping. For the samples of this work the large difference in the saturation DW velocities is related mainly to the very different magnetisations. However, in general, we predict that the maximum attainable DW velocity may be optimised by tuning the magnetisation of the magnetic layer or the DMI strength. Moreover, the measurement of v$_\mathrm{W}$ should  provide a parameter-free estimation of the DMI strength. The suppression of the Walker breakdown in samples with DMI creates a speed plateau that may ease the estimation of v$_\mathrm{W}$  in cases when the propagation field may be relatively large, hiding the linear regime. We alert that, although the plateau velocity seems to be close to v$_\mathrm{W}$ with the current parameters, this may not always be the case. The relation between v$_\mathrm{W}$ and the plateau velocity is not straightforward and deserves further study.

\section{Conclusion}
We have shown that the DMI stabilises chiral N\'{e}el walls and allows reaching very large field-driven DW velocities. The velocity saturates beyond the Walker field and no breakdown is observed,  as already found in Co/Ni/Co ultrathin films \cite{Yoshimura2015}. We demonstrate experimentally that, in agreement with the 1D model \cite{Thiaville2012}, the Walker velocity is proportional to the ratio $D/M_\mathrm{s}$, so that its measurement provides a  direct method to obtain the DMI strength.

\section{Acknowledgments}
We acknowledge the support of the Agence Nationale de la Recherche, project ANR-14-CE26-0012 (ULTRASKY). B. Fernandez and Ph. David helped in the development of the microcoils used for this work. D.S.C was supported by a CNPq Scholarship (Brazil).

\bibliographystyle{apsrev}

\begin{thebibliography}{0}
\expandafter\ifx\csname natexlab\endcsname\relax\def\natexlab#1{#1}\fi
\expandafter\ifx\csname bibnamefont\endcsname\relax
  \def\bibnamefont#1{#1}\fi
\expandafter\ifx\csname bibfnamefont\endcsname\relax
  \def\bibfnamefont#1{#1}\fi
\expandafter\ifx\csname citenamefont\endcsname\relax
  \def\citenamefont#1{#1}\fi
\expandafter\ifx\csname url\endcsname\relax
  \def\url#1{\texttt{#1}}\fi
\expandafter\ifx\csname urlprefix\endcsname\relax\def\urlprefix{URL }\fi
\providecommand{\bibinfo}[2]{#2}
\providecommand{\eprint}[2][]{\url{#2}}

\bibitem[{\citenamefont{Parkin et~al.}(2008)\citenamefont{Parkin, Hayashi, and
  Thomas}}]{Parkin2008}
\bibinfo{author}{\bibfnamefont{S.}~\bibnamefont{Parkin}},
  \bibinfo{author}{\bibfnamefont{M.}~\bibnamefont{Hayashi}}, \bibnamefont{and}
  \bibinfo{author}{\bibfnamefont{L.}~\bibnamefont{Thomas}},
  \bibinfo{journal}{Science} \textbf{\bibinfo{volume}{320}},
  \bibinfo{pages}{190} (\bibinfo{year}{2008}).

\bibitem[{\citenamefont{Moore et~al.}(2008)\citenamefont{Moore, Miron, Gaudin,
  Serret, Auffret, Rodmacq, Schuhl, Pizzini, Vogel, and Bonfim}}]{Moore2008}
\bibinfo{author}{\bibfnamefont{T.~A.} \bibnamefont{Moore}},
  \bibinfo{author}{\bibfnamefont{I.~M.} \bibnamefont{Miron}},
  \bibinfo{author}{\bibfnamefont{G.}~\bibnamefont{Gaudin}},
  \bibinfo{author}{\bibfnamefont{G.}~\bibnamefont{Serret}},
  \bibinfo{author}{\bibfnamefont{S.}~\bibnamefont{Auffret}},
  \bibinfo{author}{\bibfnamefont{B.}~\bibnamefont{Rodmacq}},
  \bibinfo{author}{\bibfnamefont{A.}~\bibnamefont{Schuhl}},
  \bibinfo{author}{\bibfnamefont{S.}~\bibnamefont{Pizzini}},
  \bibinfo{author}{\bibfnamefont{J.}~\bibnamefont{Vogel}}, \bibnamefont{and}
  \bibinfo{author}{\bibfnamefont{M.}~\bibnamefont{Bonfim}},
  \bibinfo{journal}{Appl. Phys. Lett.} \textbf{\bibinfo{volume}{93}},
  \bibinfo{eid}{262504} (\bibinfo{year}{2008}).

\bibitem[{\citenamefont{Miron et~al.}(2011)\citenamefont{Miron, Moore,
  Szambolics, Buda-Prejbeanu, Auffret, Rodmacq, Pizzini, Vogel, Bonfim, Schuhl
  et~al.}}]{Miron2011}
\bibinfo{author}{\bibfnamefont{I.~M.} \bibnamefont{Miron}},
  \bibinfo{author}{\bibfnamefont{T.~A.}~\bibnamefont{Moore}},
  \bibinfo{author}{\bibfnamefont{H.}~\bibnamefont{Szambolics}},
  \bibinfo{author}{\bibfnamefont{L.~D.} \bibnamefont{Buda-Prejbeanu}},
  \bibinfo{author}{\bibfnamefont{S.}~\bibnamefont{Auffret}},
  \bibinfo{author}{\bibfnamefont{B.}~\bibnamefont{Rodmacq}},
  \bibinfo{author}{\bibfnamefont{S.}~\bibnamefont{Pizzini}},
  \bibinfo{author}{\bibfnamefont{J.}~\bibnamefont{Vogel}},
  \bibinfo{author}{\bibfnamefont{M.}~\bibnamefont{Bonfim}},
  \bibinfo{author}{\bibfnamefont{A.}~\bibnamefont{Schuhl}}, \bibnamefont{and}
  \bibinfo{author}{\bibfnamefont{G.}~\bibnamefont{Gaudin}},
  \bibinfo{journal}{Nature Mater.}
  \textbf{\bibinfo{volume}{10}}, \bibinfo{pages}{419} (\bibinfo{year}{2011}).

\bibitem[{\citenamefont{Ryu et~al.}(2013)\citenamefont{Ryu, L., Yang, and
  Parkin}}]{Ryu2013}
\bibinfo{author}{\bibfnamefont{K.-S.} \bibnamefont{Ryu}},
  \bibinfo{author}{\bibfnamefont{L.}~\bibnamefont{Thomas}},
  \bibinfo{author}{\bibfnamefont{S.-H.} \bibnamefont{Yang}}, \bibnamefont{and}
  \bibinfo{author}{\bibfnamefont{S.}~\bibnamefont{Parkin}},
  \bibinfo{journal}{Nature Nanotech.} \textbf{\bibinfo{volume}{8}},
  \bibinfo{pages}{527} (\bibinfo{year}{2013}).

\bibitem[{\citenamefont{Emori et~al.}(2013)\citenamefont{Emori, Bauer, Ahn,
  Martinez, and Beach}}]{Emori2013}
\bibinfo{author}{\bibfnamefont{S.}~\bibnamefont{Emori}},
  \bibinfo{author}{\bibfnamefont{U.}~\bibnamefont{Bauer}},
  \bibinfo{author}{\bibfnamefont{S.-M.} \bibnamefont{Ahn}},
  \bibinfo{author}{\bibfnamefont{E.}~\bibnamefont{Martinez}}, \bibnamefont{and}
  \bibinfo{author}{\bibfnamefont{G.}~\bibnamefont{Beach}},
  \bibinfo{journal}{Nature Mater.} \textbf{\bibinfo{volume}{12}},
  \bibinfo{pages}{611} (\bibinfo{year}{2013}).

\bibitem[{\citenamefont{Thiaville et~al.}(2012)\citenamefont{Thiaville, Rohart,
  Ju{\'{e}}, Cros, and Fert}}]{Thiaville2012}
\bibinfo{author}{\bibfnamefont{A.}~\bibnamefont{Thiaville}},
  \bibinfo{author}{\bibfnamefont{S.}~\bibnamefont{Rohart}},
  \bibinfo{author}{\bibfnamefont{E.}~\bibnamefont{Ju{\'{e}}}},
  \bibinfo{author}{\bibfnamefont{V.}~\bibnamefont{Cros}}, \bibnamefont{and}
  \bibinfo{author}{\bibfnamefont{A.}~\bibnamefont{Fert}},
  \bibinfo{journal}{EPL} \textbf{\bibinfo{volume}{100}}, \bibinfo{pages}{57002}
  (\bibinfo{year}{2012}).

\bibitem[{\citenamefont{Je et~al.}(2013)\citenamefont{Je, Kim, Yoo, Min, Lee,
  and Choe}}]{Je2013}
\bibinfo{author}{\bibfnamefont{S.-G.} \bibnamefont{Je}},
  \bibinfo{author}{\bibfnamefont{D.-H.} \bibnamefont{Kim}},
  \bibinfo{author}{\bibfnamefont{S.-C.} \bibnamefont{Yoo}},
  \bibinfo{author}{\bibfnamefont{B.-C.} \bibnamefont{Min}},
  \bibinfo{author}{\bibfnamefont{K.-J.} \bibnamefont{Lee}}, \bibnamefont{and}
  \bibinfo{author}{\bibfnamefont{S.-B.} \bibnamefont{Choe}},
  \bibinfo{journal}{Phys. Rev. B} \textbf{\bibinfo{volume}{88}},
  \bibinfo{pages}{214401} (\bibinfo{year}{2013}).

\bibitem[{\citenamefont{Hrabec et~al.}(2014)\citenamefont{Hrabec, Porter,
  Wells, Benitez, Burnell, McVitie, McGrouther, Moore, and
  Marrows}}]{Hrabec2014}
\bibinfo{author}{\bibfnamefont{A.}~\bibnamefont{Hrabec}},
  \bibinfo{author}{\bibfnamefont{N.~A.} \bibnamefont{Porter}},
  \bibinfo{author}{\bibfnamefont{A.}~\bibnamefont{Wells}},
  \bibinfo{author}{\bibfnamefont{M.~J.} \bibnamefont{Benitez}},
  \bibinfo{author}{\bibfnamefont{G.}~\bibnamefont{Burnell}},
  \bibinfo{author}{\bibfnamefont{S.}~\bibnamefont{McVitie}},
  \bibinfo{author}{\bibfnamefont{D.}~\bibnamefont{McGrouther}},
  \bibinfo{author}{\bibfnamefont{T.~A.} \bibnamefont{Moore}}, \bibnamefont{and}
  \bibinfo{author}{\bibfnamefont{C.~H.} \bibnamefont{Marrows}},
  \bibinfo{journal}{Phys. Rev. B} \textbf{\bibinfo{volume}{90}},
  \bibinfo{pages}{020402} (\bibinfo{year}{2014}).

\bibitem[{\citenamefont{Lavrijsen et~al.}(2015)\citenamefont{Lavrijsen,
  Hartmann, van~den Brink, Yin, Barcones, Duine, Verheijen, Swagten, and
  Koopmans}}]{Lavrijsen2015}
\bibinfo{author}{\bibfnamefont{R.}~\bibnamefont{Lavrijsen}},
  \bibinfo{author}{\bibfnamefont{D.~M.~F.} \bibnamefont{Hartmann}},
  \bibinfo{author}{\bibfnamefont{A.}~\bibnamefont{van~den Brink}},
  \bibinfo{author}{\bibfnamefont{Y.}~\bibnamefont{Yin}},
  \bibinfo{author}{\bibfnamefont{B.}~\bibnamefont{Barcones}},
  \bibinfo{author}{\bibfnamefont{R.~A.} \bibnamefont{Duine}},
  \bibinfo{author}{\bibfnamefont{M.~A.} \bibnamefont{Verheijen}},
  \bibinfo{author}{\bibfnamefont{H.~J.~M.} \bibnamefont{Swagten}},
  \bibnamefont{and} \bibinfo{author}{\bibfnamefont{B.}~\bibnamefont{Koopmans}},
  \bibinfo{journal}{Phys. Rev. B} \textbf{\bibinfo{volume}{91}},
  \bibinfo{pages}{104414} (\bibinfo{year}{2015}).

\bibitem[{\citenamefont{Vanatka et~al.}(2015)\citenamefont{Vanatka,
  Rojas-Sanchez, Vogel, Bonfim, Thiaville, and Pizzini}}]{Vanatka2015}
\bibinfo{author}{\bibfnamefont{M.}~\bibnamefont{Vanatka}},
  \bibinfo{author}{\bibfnamefont{J.-C.} \bibnamefont{Rojas-Sanchez}},
  \bibinfo{author}{\bibfnamefont{J.}~\bibnamefont{Vogel}},
  \bibinfo{author}{\bibfnamefont{M.}~\bibnamefont{Bonfim}},
  \bibinfo{author}{\bibfnamefont{A.}~\bibnamefont{Thiaville}},
  \bibnamefont{and} \bibinfo{author}{\bibfnamefont{S.}~\bibnamefont{Pizzini}},
  \bibinfo{journal}{J. Phys.: Condens. Matter} \textbf{\bibinfo{volume}{27}},
  \bibinfo{pages}{32002} (\bibinfo{year}{2015}).

\bibitem[{\citenamefont{Yoshimura et~al.}(2015)\citenamefont{Yoshimura, Kim,
  Taniguchi, Tono, Ueda, Hiramatsu, Moriyama, Yamada, Nakatani, and
  Ono}}]{Yoshimura2015}
\bibinfo{author}{\bibfnamefont{Y.}~\bibnamefont{Yoshimura}},
  \bibinfo{author}{\bibfnamefont{K.-J.} \bibnamefont{Kim}},
  \bibinfo{author}{\bibfnamefont{T.}~\bibnamefont{Taniguchi}},
  \bibinfo{author}{\bibfnamefont{T.}~\bibnamefont{Tono}},
  \bibinfo{author}{\bibfnamefont{K.}~\bibnamefont{Ueda}},
  \bibinfo{author}{\bibfnamefont{R.}~\bibnamefont{Hiramatsu}},
  \bibinfo{author}{\bibfnamefont{T.}~\bibnamefont{Moriyama}},
  \bibinfo{author}{\bibfnamefont{N.}~\bibnamefont{Yamada}},
  \bibinfo{author}{\bibfnamefont{Y.}~\bibnamefont{Nakatani}}, \bibnamefont{and}
  \bibinfo{author}{\bibfnamefont{T.}~\bibnamefont{Ono}}, \bibinfo{journal}{Nat.
  Phys.}  (\bibinfo{year}{2015}).

\bibitem[{\citenamefont{Yamada and Nakatani}(2015)}]{Yamada2015}
\bibinfo{author}{\bibfnamefont{K.}~\bibnamefont{Yamada}} \bibnamefont{and}
  \bibinfo{author}{\bibfnamefont{Y.}~\bibnamefont{Nakatani}},
  \bibinfo{journal}{Appl. Phys. Express} \textbf{\bibinfo{volume}{8}},
  \bibinfo{pages}{093004} (\bibinfo{year}{2015}).

\bibitem[{\citenamefont{Ju\'e et~al.}(2016)\citenamefont{Ju\'e, Thiaville,
  Pizzini, Miltat, Sampaio, Buda-Prejbeanu, Rohart, Vogel, Bonfim, Boulle
  et~al.}}]{Jue2016}
\bibinfo{author}{\bibfnamefont{E.}~\bibnamefont{Ju\'e}},
  \bibinfo{author}{\bibfnamefont{A.}~\bibnamefont{Thiaville}},
  \bibinfo{author}{\bibfnamefont{S.}~\bibnamefont{Pizzini}},
  \bibinfo{author}{\bibfnamefont{J.}~\bibnamefont{Miltat}},
  \bibinfo{author}{\bibfnamefont{J.}~\bibnamefont{Sampaio}},
  \bibinfo{author}{\bibfnamefont{L.~D.} \bibnamefont{Buda-Prejbeanu}},
  \bibinfo{author}{\bibfnamefont{S.}~\bibnamefont{Rohart}},
  \bibinfo{author}{\bibfnamefont{J.}~\bibnamefont{Vogel}},
  \bibinfo{author}{\bibfnamefont{M.}~\bibnamefont{Bonfim}},
  \bibinfo{author}{\bibfnamefont{O.}~\bibnamefont{Boulle}},
  \bibinfo{author}{\bibfnamefont{S.}~\bibnamefont{Auffret}},
  \bibinfo{author}{\bibfnamefont{I.M.}~\bibnamefont{Miron}}, \bibnamefont{and}
  \bibinfo{author}{\bibfnamefont{G.}~\bibnamefont{Gaudin}},
  \bibinfo{journal}{Phys. Rev. B}
  \textbf{\bibinfo{volume}{93}}, \bibinfo{pages}{014403}
  (\bibinfo{year}{2016}).

\bibitem[{\citenamefont{Metaxas et~al.}(2007)\citenamefont{Metaxas, Jamet,
  Mougin, Cormier, Ferr\'e, Baltz, Rodmacq, Dieny, and Stamps}}]{Metaxas2007}
\bibinfo{author}{\bibfnamefont{P.~J.} \bibnamefont{Metaxas}},
  \bibinfo{author}{\bibfnamefont{J.~P.} \bibnamefont{Jamet}},
  \bibinfo{author}{\bibfnamefont{A.}~\bibnamefont{Mougin}},
  \bibinfo{author}{\bibfnamefont{M.}~\bibnamefont{Cormier}},
  \bibinfo{author}{\bibfnamefont{J.}~\bibnamefont{Ferr\'e}},
  \bibinfo{author}{\bibfnamefont{V.}~\bibnamefont{Baltz}},
  \bibinfo{author}{\bibfnamefont{B.}~\bibnamefont{Rodmacq}},
  \bibinfo{author}{\bibfnamefont{B.}~\bibnamefont{Dieny}}, \bibnamefont{and}
  \bibinfo{author}{\bibfnamefont{R.~L.} \bibnamefont{Stamps}},
  \bibinfo{journal}{Phys. Rev. Lett.} \textbf{\bibinfo{volume}{99}},
  \bibinfo{pages}{217208} (\bibinfo{year}{2007}).

\bibitem[{\citenamefont{Dzyaloshinskii}(1957)}]{Dzyaloshinskii1957}
\bibinfo{author}{\bibfnamefont{I.~E.} \bibnamefont{Dzyaloshinskii}},
  \bibinfo{journal}{Sov. Phys. JETP} \textbf{\bibinfo{volume}{5}},
  \bibinfo{pages}{1259} (\bibinfo{year}{1957}).

\bibitem[{\citenamefont{Moriya}(1960)}]{Moriya1960}
\bibinfo{author}{\bibfnamefont{T.}~\bibnamefont{Moriya}},
  \bibinfo{journal}{Phys. Rev.} \textbf{\bibinfo{volume}{120}},
  \bibinfo{pages}{91} (\bibinfo{year}{1960}).

\bibitem[{\citenamefont{Fert}(1990)}]{Fert1990}
\bibinfo{author}{\bibfnamefont{A.}~\bibnamefont{Fert}},
  \bibinfo{journal}{Mater. Sci. Forum} \textbf{\bibinfo{volume}{59}},
  \bibinfo{pages}{439} (\bibinfo{year}{1990}).

\bibitem[{\citenamefont{Pizzini et~al.}(2014)\citenamefont{Pizzini, Vogel,
  Rohart, Buda-Prejbeanu, Ju\'{e}, Boulle, Miron, Safeer, Auffret, Gaudin
  et~al.}}]{Pizzini2014}
\bibinfo{author}{\bibfnamefont{S.}~\bibnamefont{Pizzini}},
  \bibinfo{author}{\bibfnamefont{J.}~\bibnamefont{Vogel}},
  \bibinfo{author}{\bibfnamefont{S.}~\bibnamefont{Rohart}},
  \bibinfo{author}{\bibfnamefont{L.}~\bibnamefont{Buda-Prejbeanu}},
  \bibinfo{author}{\bibfnamefont{E.}~\bibnamefont{Ju\'{e}}},
  \bibinfo{author}{\bibfnamefont{O.}~\bibnamefont{Boulle}},
  \bibinfo{author}{\bibfnamefont{I.}~\bibnamefont{Miron}},
  \bibinfo{author}{\bibfnamefont{C.}~\bibnamefont{Safeer}},
  \bibinfo{author}{\bibfnamefont{S.}~\bibnamefont{Auffret}},
  \bibinfo{author}{\bibfnamefont{G.}~\bibnamefont{Gaudin}}, \bibnamefont{and}
  \bibinfo{author}{\bibfnamefont{A.}~\bibnamefont{Thiaville}},
  \bibinfo{journal}{Phys. Rev. Lett.}
  \textbf{\bibinfo{volume}{113}}, \bibinfo{pages}{047203}
  (\bibinfo{year}{2014}).

\bibitem[{\citenamefont{Belmeguenai et~al.}(2015)\citenamefont{Belmeguenai,
  Adam, Roussign\'{e}, Eimer, Devolder, Kim, Cherif, Stashkevich, and
  Thiaville}}]{Belmeguenai2015}
\bibinfo{author}{\bibfnamefont{M.}~\bibnamefont{Belmeguenai}},
  \bibinfo{author}{\bibfnamefont{J.-P.} \bibnamefont{Adam}},
  \bibinfo{author}{\bibfnamefont{Y.}~\bibnamefont{Roussign\'{e}}},
  \bibinfo{author}{\bibfnamefont{S.}~\bibnamefont{Eimer}},
  \bibinfo{author}{\bibfnamefont{T.}~\bibnamefont{Devolder}},
  \bibinfo{author}{\bibfnamefont{J.-V.} \bibnamefont{Kim}},
  \bibinfo{author}{\bibfnamefont{S.}~\bibnamefont{Cherif}},
  \bibinfo{author}{\bibfnamefont{A.}~\bibnamefont{Stashkevich}},
  \bibnamefont{and}
  \bibinfo{author}{\bibfnamefont{A.}~\bibnamefont{Thiaville}},
  \bibinfo{journal}{Phys. Rev. B} \textbf{\bibinfo{volume}{91}},
  \bibinfo{pages}{180405(R)} (\bibinfo{year}{2015}).

\bibitem[{\citenamefont{Ju\'e et~al.}(2015)\citenamefont{Ju\'e, Safeer,
  Drouard, Lopez, Balint, Buda-Prejbeanu, Boulle, Auffret, Schuhl, Manchon
  et~al.}}]{Jue2015a}
\bibinfo{author}{\bibfnamefont{E.}~\bibnamefont{Ju\'e}},
  \bibinfo{author}{\bibfnamefont{C.}~\bibnamefont{Safeer}},
  \bibinfo{author}{\bibfnamefont{M.}~\bibnamefont{Drouard}},
  \bibinfo{author}{\bibfnamefont{A.}~\bibnamefont{Lopez}},
  \bibinfo{author}{\bibfnamefont{P.}~\bibnamefont{Balint}},
  \bibinfo{author}{\bibfnamefont{L.}~\bibnamefont{Buda-Prejbeanu}},
  \bibinfo{author}{\bibfnamefont{O.}~\bibnamefont{Boulle}},
  \bibinfo{author}{\bibfnamefont{S.}~\bibnamefont{Auffret}},
  \bibinfo{author}{\bibfnamefont{A.}~\bibnamefont{Schuhl}},
  \bibinfo{author}{\bibfnamefont{A.}~\bibnamefont{Manchon}},
  \bibinfo{author}{\bibfnamefont{I.M.}~\bibnamefont{Miron}}, \bibnamefont{and}
  \bibinfo{author}{\bibfnamefont{G.}~\bibnamefont{Gaudin}},
  \bibinfo{journal}{Nature Mater.}
  (\bibinfo{year}{2015}).

\bibitem[{\citenamefont{Tetienne et~al.}(2014)\citenamefont{Tetienne, Hingant,
  Mart\`{\i}nez, Rohart, Thiaville, Herrera~Diez, Garcia, Adam, Kim, Roch
  et~al.}}]{Tetienne2014}
\bibinfo{author}{\bibfnamefont{J.}~\bibnamefont{Tetienne}},
  \bibinfo{author}{\bibfnamefont{T.}~\bibnamefont{Hingant}},
  \bibinfo{author}{\bibfnamefont{L.}~\bibnamefont{Mart\`{\i}nez}},
  \bibinfo{author}{\bibfnamefont{S.}~\bibnamefont{Rohart}},
  \bibinfo{author}{\bibfnamefont{A.}~\bibnamefont{Thiaville}},
  \bibinfo{author}{\bibfnamefont{L.}~\bibnamefont{Herrera~Diez}},
  \bibinfo{author}{\bibfnamefont{K.}~\bibnamefont{Garcia}},
  \bibinfo{author}{\bibfnamefont{J.-P.} \bibnamefont{Adam}},
  \bibinfo{author}{\bibfnamefont{J.-V.} \bibnamefont{Kim}},
  \bibinfo{author}{\bibfnamefont{J.-F.} \bibnamefont{Roch}},
  \bibinfo{author}{\bibfnamefont{I.M.} \bibnamefont{Miron}},
  \bibinfo{author}{\bibfnamefont{G.} \bibnamefont{Gaudin}},
  \bibinfo{author}{\bibfnamefont{L.} \bibnamefont{Vila}},
  \bibinfo{author}{\bibfnamefont{B.} \bibnamefont{Ocker}},
  \bibinfo{author}{\bibfnamefont{D.} \bibnamefont{Ravelosona}}, \bibnamefont{and}
  \bibinfo{author}{\bibfnamefont{V.} \bibnamefont{Jacques}},
  \bibinfo{journal}{Nature Commun.}
  \textbf{\bibinfo{volume}{5}}, \bibinfo{pages}{6733} (\bibinfo{year}{2014}).

\bibitem[{\citenamefont{Vansteenkiste et~al.}(2014)\citenamefont{Vansteenkiste,
  Leliaert, Dvornik, Helsen, Garcia-Sanchez, and Van~Waeyenberge}}]{Mumax3}
\bibinfo{author}{\bibfnamefont{A.}~\bibnamefont{Vansteenkiste}},
  \bibinfo{author}{\bibfnamefont{J.}~\bibnamefont{Leliaert}},
  \bibinfo{author}{\bibfnamefont{M.}~\bibnamefont{Dvornik}},
  \bibinfo{author}{\bibfnamefont{M.}~\bibnamefont{Helsen}},
  \bibinfo{author}{\bibfnamefont{F.}~\bibnamefont{Garcia-Sanchez}},
  \bibnamefont{and}
  \bibinfo{author}{\bibfnamefont{B.}~\bibnamefont{Van~Waeyenberge}},
  \bibinfo{journal}{AIP Advances} \textbf{\bibinfo{volume}{4}},
  \bibinfo{eid}{107133} (\bibinfo{year}{2014}).

\bibitem[{\citenamefont{Thole et~al.}(1992)\citenamefont{Thole, Carra, Sette,
  and van~der Laan}}]{Thole1992}
\bibinfo{author}{\bibfnamefont{B.~T.} \bibnamefont{Thole}},
  \bibinfo{author}{\bibfnamefont{P.}~\bibnamefont{Carra}},
  \bibinfo{author}{\bibfnamefont{F.}~\bibnamefont{Sette}}, \bibnamefont{and}
  \bibinfo{author}{\bibfnamefont{G.}~\bibnamefont{van~der Laan}},
  \bibinfo{journal}{Phys. Rev. Lett.} \textbf{\bibinfo{volume}{68}},
  \bibinfo{pages}{1943} (\bibinfo{year}{1992}).

\bibitem[{\citenamefont{Carra et~al.}(1993)\citenamefont{Carra, Thole,
  Altarelli, and Wang}}]{Carra1993}
\bibinfo{author}{\bibfnamefont{P.}~\bibnamefont{Carra}},
  \bibinfo{author}{\bibfnamefont{B.~T.} \bibnamefont{Thole}},
  \bibinfo{author}{\bibfnamefont{M.}~\bibnamefont{Altarelli}},
  \bibnamefont{and} \bibinfo{author}{\bibfnamefont{X.}~\bibnamefont{Wang}},
  \bibinfo{journal}{Phys. Rev. Lett.} \textbf{\bibinfo{volume}{70}},
  \bibinfo{pages}{694} (\bibinfo{year}{1993}).

\bibitem[{\citenamefont{L\'opez-Flores
  et~al.}(2013)\citenamefont{L\'opez-Flores, Bergeard, Halt\'e, Stamm, Pontius,
  Hehn, Otero, Beaurepaire, and Boeglin}}]{Boeglin2013}
\bibinfo{author}{\bibfnamefont{V.}~\bibnamefont{L\'opez-Flores}},
  \bibinfo{author}{\bibfnamefont{N.}~\bibnamefont{Bergeard}},
  \bibinfo{author}{\bibfnamefont{V.}~\bibnamefont{Halt\'e}},
  \bibinfo{author}{\bibfnamefont{C.}~\bibnamefont{Stamm}},
  \bibinfo{author}{\bibfnamefont{N.}~\bibnamefont{Pontius}},
  \bibinfo{author}{\bibfnamefont{M.}~\bibnamefont{Hehn}},
  \bibinfo{author}{\bibfnamefont{E.}~\bibnamefont{Otero}},
  \bibinfo{author}{\bibfnamefont{E.}~\bibnamefont{Beaurepaire}},
  \bibnamefont{and} \bibinfo{author}{\bibfnamefont{C.}~\bibnamefont{Boeglin}},
  \bibinfo{journal}{Phys. Rev. B} \textbf{\bibinfo{volume}{87}},
  \bibinfo{pages}{214412} (\bibinfo{year}{2013}).

\end{thebibliography}

\end{document}